\documentclass[journal]{IEEEtran}

\usepackage{amsmath,amssymb}
\usepackage{graphicx}
\usepackage{booktabs}
\usepackage{xcolor}
\usepackage[hidelinks]{hyperref}
\usepackage{cite}
\usepackage{array}

\newcommand{\ie}{\mathrm{IE}}
\newcommand{\nnkr}{\mathrm{NNKR}}
\newcommand{\ce}{\mathrm{CE}}
\newcommand{\skr}{R_\mathrm{sec}}
\newcommand{\qber}{Q}
\newcommand{\yes}{\checkmark}
\newcommand{\nope}{--}

\begin{document}

\title{Benchmarking Quantum Key Distribution Hardware Across Operating Conditions}

\author{Jerry~Horgan,~\IEEEmembership{Senior~Member,~IEEE,}~Indrakshi~Dey,~\IEEEmembership{Senior~Member,~IEEE,}~Hajar~Bennouri,~and~Daniel~C.~Kilper,~\IEEEmembership{Senior~Member,~IEEE}
\thanks{J.~Horgan (ORCID: 0000-0001-5783-6077), H.~Bennouri (ORCID: 0000-0003-2358-390X) and D.~C.~Kilper (ORCID: 0000-0003-3542-5335) are with the School of Electrical Engineering, CONNECT Centre, Trinity College Dublin, Dublin~2, Ireland (e-mail: horganj3@tcd.ie).}
\thanks{J.~Horgan, and I.~Dey (ORCID:0000-0001-9669-6417) are with the Department of Computing and Mathematics, South East Technological University, Ireland}
\thanks{This work was supported by Research Ireland under Grants 21/US-C2C/3750 and 13/RC/2077\_P2.}
\thanks{\emph{This work is under review at IEEE Journal of Lightwave Technology.}}}

\maketitle

\begin{abstract}
Quantum key distribution (QKD) is entering national fibre networks, but published secret key rates (SKRs) are measured over different distances, wavelengths, classical traffic loads and clock rates, so systems are hard to compare. We write the SKR as the product of the transmitter clock rate, the channel transmittance and the secret key per received pulse. This third factor, the noise-normalised key rate (NNKR), needs four commonly reported parameters and depends mainly on the equipment wherever the SKR is proportional to transmittance. A companion metric, the implementation efficiency (IE),  compares the SKR with the ideal key rate of its channel, including classical-traffic noise. For fixed loss-limited equipment, NNKR is independent of channel loss, whereas the fraction of the repeaterless bound varies 1.6-fold over 2--25~dB. An analytic model relates the flat loss range of NNKR to the detector noise count rate and dead time. We apply both metrics to ten published coexistence experiments, and NNKR to 42,115 measurements from four systems of two vendors and four high-rate laboratory systems. For one production system, NNKR has a 6.9\% coefficient of variation over 17~dB of attenuation and agrees within 0.7~dB with three of four published deployments of that platform. Under coexistence, that platform's NNKR and IE stay within 0.5~dB of their standalone values. In decibels, NNKR is a sum of component penalties, as in an optical link budget, and for one laboratory system the budget from its published parameters and measured secure fraction closes within 0.3~dB.
\end{abstract}


\section{Introduction}
\label{sec:intro}

\IEEEPARstart{Q}{uantum} key distribution (QKD) allows two parties to establish shared secret
keys with security proofs based on quantum mechanics under stated device assumptions~\cite{Pirandola2020_Advances}. The technology is moving from point-to-point demonstrations into national infrastructure: all 27 European Union member states have signed the declaration of the European Quantum Communication Infrastructure (EuroQCI), and national quantum communication networks have been under construction since the first implementation phase began in 2023~\cite{EuroQCI}. Many of these networks are planned on installed fibre that carries classical traffic. Coexistence of quantum and classical signals has been demonstrated across protocols, fibre types, wavelength bands~\cite{Patel2014,Dynes2016,Wang2017,Bacco2019,Eriksson2019,Valivarthi2019,HuguesSalas2020,Alia2022,Pistoia2023,Gavignet2024_JLT} and classical capacities up to 18~Tb/s~\cite{Eriksson2019}.

As deployment grows, operators, vendors and researchers share one practical question: on the same channel, which implementation extracts more secret key? Every experiment reports its secret key rate (SKR) at its own distance, wavelength, launch power and clock rate, so each reported value combines the quality of the hardware with the conditions of the channel. The SKR remains the figure that matters for a given link; the difficulty is that each published value is measured under conditions its authors chose. Plots of SKR against distance describe individual links, and the bound of Pirandola, Laurenza, Ottaviani and Banchi (PLOB)~\cite{Pirandola2017_PLOB} gives the maximum secret key per channel use of a lossy repeaterless channel. The co-propagation efficiency (CE) of Gavignet \textit{et al.}~\cite{Gavignet2024_JLT} measures coexistence capacity, and Bebrov's efficiency~\cite{Bebrov2024} counts the resources a protocol consumes. Recommendation Y.3807 of the Telecommunication Standardization Sector of the International Telecommunication Union (ITU-T)~\cite{ITUT_Y3807_2022} lists quality-of-service parameters.

This paper proposes a hardware metric for QKD and tests it on published and testbed data. We write the SKR as the product of the clock rate, the transmittance and a hardware term, the noise-normalised key rate (NNKR), which in decibels is approximately a sum of source, receiver, detector and key-processing factors. At high loss NNKR equals the PLOB fraction, the secret key per pulse divided by the PLOB capacity, divided by $\ln 2$; unlike the PLOB fraction, it does not change with loss for fixed loss-limited hardware, and it complements CE, which by design scales with fibre length and classical launch power. We also define the implementation efficiency (IE), whose reference is the ideal key of the same lossy, noisy channel computed from the closed-form coexistence limits of Zischler \textit{et al.}~\cite{Zischler2026}, and relate the two metrics for BB84. To our knowledge, a per-pulse hardware metric of this kind has not previously been tested for loss invariance on a multi-vendor testbed, or used to place coexistence experiments, production systems and laboratory records on one scale.

\begin{figure*}[htbp]
\centering
\includegraphics[width=\textwidth]{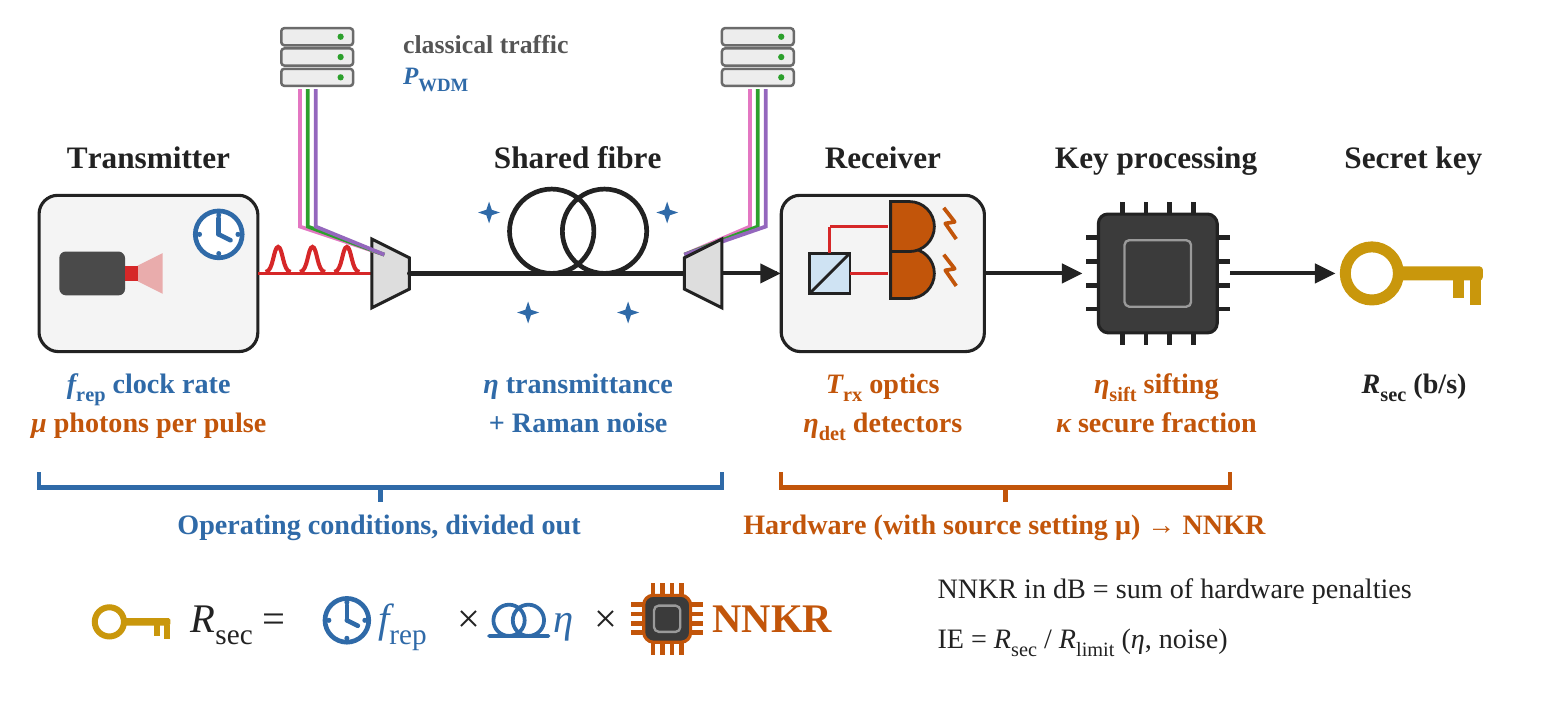}
\caption{Conceptual overview: The transmitter emits $f_\mathrm{rep}$ pulses per second (clock rate), each with mean photon number $\mu$; the shared fibre transmits a fraction $\eta$ of them (transmittance) and adds Raman noise from the classical traffic of total launch power $P_\mathrm{WDM}$; the receiver optics (transmission $T_\mathrm{rx}$), the single-photon detectors (efficiency $\eta_\mathrm{det}$) and key processing (sifting efficiency $\eta_\mathrm{sift}$, secure fraction $\kappa$) turn each arriving pulse into secret key at the secret key rate $R_\mathrm{sec}$. The clock rate and the transmittance are operating conditions (blue) and are divided out; the noise-normalised key rate NNKR collects the hardware factors (orange) and, in decibels, is their sum. The implementation efficiency IE compares $R_\mathrm{sec}$ with the ideal key rate $R_\mathrm{limit}$ of the same lossy, noisy channel. WDM: wavelength-division multiplexing.}
\label{fig:concept}
\end{figure*}

This work makes four contributions. First, it introduces NNKR, which uses four commonly reported parameters and decomposes into source, receiver, detector and key-processing factors. An analytic key-rate model gives the hardware constant in closed form and the edges of the loss range over which NNKR is flat, set by the product of detector noise count rate and dead time. Second, it introduces IE and shows that for the Bennett--Brassard 1984 (BB84) protocol, IE/NNKR $= 2/[1-2H_2(\qber_\mathrm{model})]$,  where $H_2$ is the binary entropy function and $\qber_\mathrm{model}$ the error rate that the modelled channel noise causes, so that the ratio shows how much the modelled noise changes the reference; it also
derives the noise sensitivity of IE and the error propagation of both metrics. Third, it tests the metrics on ten published coexistence experiments, on 42{,}115 measurements from four QKD systems of two vendors and on four high-rate laboratory records, with five consistency checks, and it tests whether either metric degrades when a production system operates in a coexistence environment. Fourth, it places production systems and laboratory records on one NNKR scale, closes the budget of one record system within 0.3~dB using its measured secure fraction, finds that two recent records were measured where detector saturation or afterpulsing lowers NNKR, and compares NNKR with CE and the PLOB fraction (Table~\ref{tab:metric_comparison}). 

Section~\ref{sec:concept} presents the concept, Section~\ref{sec:metrics} defines the metrics, Section~\ref{sec:math} develops their mathematical analysis, Section~\ref{sec:method} describes the data sets and the consistency checks, Section~\ref{sec:results} presents the results with their physical interpretation and the comparison with the state of the art, and Section~\ref{sec:conclusion} concludes.

\section{Concept}
\label{sec:concept}

Fig.~\ref{fig:concept} illustrates the conceptual overview of a QKD system. A QKD link has four stages. The transmitter emits pulses at a clock rate of $f_\mathrm{rep}$ pulses per second, each with a mean photon number $\mu$. The fibre transmits a fraction $\eta = 10^{-\alpha L/10}$ of the light, where $\alpha$ is the attenuation in dB/km and $L$ the length, and in a shared fibre it also adds noise photons generated by the classical wavelength-division-multiplexed (WDM) channels, whose total launch power we denote $P_\mathrm{WDM}$. The dominant noise mechanism is spontaneous Raman scattering (SpRS) \cite{Peters2009,Eraerds2010}, in which molecular vibrations of the glass shift a small fraction of the classical light into the quantum channel. The receiver passes a fraction $T_\mathrm{rx}$ of the arriving light to single-photon detectors of efficiency $\eta_\mathrm{det}$, and key processing converts the detections into secret key through sifting (keeping the detections in which sender and receiver used the same measurement basis), error correction and privacy amplification (shortening the key to remove any information an eavesdropper may hold). The secret key rate $\skr$ therefore factorises as
\begin{equation}
\skr = f_\mathrm{rep}\cdot\eta\cdot\nnkr ,
\label{eq:factor}
\end{equation}
where the first factor states how fast the transmitter emits pulses, the second how many pulses survive the channel, and the third, the noise-normalised key rate, how much secret key the hardware extracts from each pulse that reaches the receiver. The transmittance changes from one deployment to the next, and the clock rate is a property of the transmitter. NNKR depends on the source, receiver, detectors and key processing. For decoy-state BB84, in which the transmitter varies the pulse intensity at random so that the contribution of single photons can be bounded~\cite{Pirandola2020_Advances}, NNKR expands into a product of physical factors,
\begin{equation}
\nnkr \approx \mu\,T_\mathrm{rx}\,\eta_\mathrm{det}\,\eta_\mathrm{sift}\,\kappa ,
\label{eq:budget}
\end{equation}
where $\eta_\mathrm{sift}$ is the fraction of detections kept after basis reconciliation and $\kappa$ is the secure fraction, the number of secret bits per sifted bit after error correction and privacy amplification~\cite{Yuan2018}; $\kappa$ falls as the quantum bit error rate (QBER), the fraction of sifted bits in error, rises. In decibels (\ref{eq:budget}) becomes a sum, so NNKR reads as a penalty budget in the same way as a classical optical link budget. Ideal single-photon BB84 with standard sifting corresponds to $\mu = T_\mathrm{rx} = \eta_\mathrm{det} = \kappa = 1$ and $\eta_\mathrm{sift} = 1/2$, that is $-3.0$~dB, and an ideal decoy-state system with weak coherent pulses and biased basis choice reaches $\mu e^{-\mu} \approx 0.30$ ($-5.2$~dB) at a typical $\mu = 0.5$, with $\eta_\mathrm{sift}$ close to 1~\cite{Pirandola2020_Advances}. The implementation efficiency completes the picture by moving the channel noise into the reference. It divides the SKR by the key that ideal equipment would extract from the same lossy, noisy channel.

\section{Metrics}
\label{sec:metrics}

\begin{table*}[t]
\centering
\caption{Ten QKD Coexistence Experiments Used in This Study}
\label{tab:experiments}
\small\setlength{\tabcolsep}{4.2pt}
\begin{tabular*}{\textwidth}{@{\extracolsep{\fill}}llcccccccccr@{}}
\toprule
\textbf{Experiment} & \textbf{Protocol} & \textbf{$\lambda_Q$} & \textbf{Fibre} & \textbf{$L$} & \textbf{$P_\mathrm{WDM}$} & \textbf{SKR} & \textbf{QBER} & \textbf{CE} & \textbf{NNKR} & \textbf{IE} & \textbf{IE/NNKR} \\
 & & (nm) & & (km) & (dBm) & (b/s) & (\%) & & & & \\
\midrule
Patel'14~\cite{Patel2014} & BB84 & 1550 & DSF & 50 & $-$14 & 1.17M & 2.0 & $2.3\!\times\!10^6$ & 0.0117 & 0.0234 & 2.00 \\
Dynes'16~\cite{Dynes2016} & BB84 & 1547 & SSMF & 50 & $-$25.5 & 1.2M & 2.7 & $1.7\!\times\!10^5$ & 0.0107 & 0.0214 & 2.00 \\
Wang'17~\cite{Wang2017} & BB84 & 1310 & SSMF & 50 & 4 & 18.7k & 1.5 & $2.4\!\times\!10^6$ & 0.00134 & 0.00271 & 2.03 \\
Bacco'19~\cite{Bacco2019} & BB84 & 1550 & 37-MCF & 7.9 & $-$34 & 1.7M$^a$ & 0.8 & $5.3\!\times\!10^3$ & 0.00467 & 0.00934 & 2.00 \\
Eriksson'19~\cite{Eriksson2019} & CV-QKD & 1550 & SSMF & 10 & 12.9 & 28.9k & --- & $5.6\!\times\!10^6$ & 0.00458 & 0.0035$^b$ & 0.75 \\
Valivarthi'19~\cite{Valivarthi2019} & MDI & 1532 & SSMF & 40$^e$ & $-$27 & 167$^e$ & --- & 13.4 & $5.3\!\times\!10^{-5}$ & ---$^b$ & --- \\
Hugues-Salas'20~\cite{HuguesSalas2020} & BB84 & 1552 & 7-MCF & 1 & $-$9.5 & 73 & 6.3 & 8.2 & $3.0\!\times\!10^{-7}$ & ${\geq}6.0\!\times\!10^{-7\,c}$ & $\geq$2.00 \\
Alia'22~\cite{Alia2022} & COW & 1548 & HC-NANF & 2 & 0 & 2.3k & 2.8 & $4.6\!\times\!10^3$ & $6.7\!\times\!10^{-6}$ & ---$^b$ & --- \\
Pistoia'23~\cite{Pistoia2023} & BB84 & 1313 & SSMF & 70 & 0$^d$ & 66k & 4.0 & $4.6\!\times\!10^6$ & 0.0115 & 0.0232 & 2.02 \\
Gavignet'24~\cite{Gavignet2024_JLT} & BB84 & 1310 & SSMF & 50 & 16.8 & 106k & 5.4 & $2.5\!\times\!10^8$ & 0.00422 & 0.0094 & 2.23 \\
\bottomrule
\multicolumn{12}{@{}p{\textwidth}@{}}{\footnotesize $\lambda_Q$: quantum-channel wavelength; $L$: fibre length; $P_\mathrm{WDM}$: total classical launch power; QBER: quantum bit error rate; CE in b/s$\cdot$mW$\cdot$km. Protocols: BB84, decoy-state BB84; CV-QKD, continuous-variable QKD; MDI, measurement-device-independent QKD; COW, coherent one-way. Fibres: DSF, dispersion-shifted fibre; SSMF, standard single-mode fibre; $n$-MCF, multicore fibre with $n$ cores; HC-NANF, hollow-core nested antiresonant nodeless fibre. $^a$Per core (62.8~Mb/s over 37 cores); all metrics per core. $^b$Against the Gaussian-modulated limit~\cite{Zischler2026}, a conservative reference for this four-state protocol. The coexistence limits of~\cite{Zischler2026} cover BB84 and Gaussian-modulated CV-QKD; no equivalent limit is available for COW or MDI-QKD, so only NNKR is given for these entries. $^c$Noise-free bound $2\cdot$NNKR from (\ref{eq:ratio}), evaluated with the same $f_\mathrm{rep}$ and $\eta$ as NNKR; modelled noise raises IE above this bound. $^d$Launch power of the 800-Gb/s channel under test~\cite{Pistoia2023}; the total power of the 10-channel configuration is not reported. $^e$Secret key per clock cycle with one classical channel (Table~C2 of~\cite{Valivarthi2019}) at the 20~MHz clock rate; $L$ is the total Alice--Bob length.}
\end{tabular*}
\end{table*}

\subsection{Noise-Normalised Key Rate}
\label{sec:nnkr}
Rearranging (\ref{eq:factor}) defines the noise-normalised key rate,
\begin{equation}
\nnkr = \frac{\skr}{f_\mathrm{rep}\,\eta}, \qquad \eta = 10^{-\alpha L/10},
\label{eq:nnkr}
\end{equation}
where $\eta$ is the fibre transmittance. Multiplexer, filter and connector losses belong to the setup and count as hardware in NNKR. NNKR is the secret key per transmitted pulse divided by the probability that a pulse survives the channel, and it uses four routinely reported parameters: $\skr$, $f_\mathrm{rep}$, $\alpha$ and $L$. NNKR stays constant wherever the SKR scales linearly with $\eta$, which is the loss-limited regime of prepare-and-measure discrete-variable QKD (DV-QKD) and of measurement-device-independent QKD (MDI-QKD), in which both users send photons to a central measuring node. This scaling breaks down at both ends of the loss range. At low loss, detector saturation, dead time or key-processing throughput can set the SKR, so NNKR rises as $\eta$ falls. Near the reach limit, dark counts, afterpulses and, in shared fibre, Raman photons approach the signal detection rate, the error rate rises and NNKR rolls off. NNKR is also not constant for protocols with a different rate–loss scaling, such as twin-field QKD, whose key scales as $\sqrt{\eta}$~\cite{Pirandola2020_Advances}. The range over which NNKR is flat therefore identifies where a system is limited by the channel, and Section~\ref{sec:method} defines a criterion for this range. Table~\ref{tab:experiments} lists NNKR for ten coexistence experiments.

\subsection{Implementation Efficiency}
\label{sec:iedef}
IE divides the SKR by the asymptotic key rate of an ideal transmitter and receiver on the same channel,
\begin{equation}
\ie = \frac{\skr}{R_\mathrm{limit}}, \qquad R_\mathrm{limit} = r\,f_\mathrm{rep},
\label{eq:ie}
\end{equation}
where $r$ is the ideal key per transmitted symbol at the same loss and coexistence noise. For BB84~\cite{Zischler2026},
\begin{align}
r_\mathrm{BB84} &= \tfrac{\eta}{2}\left[1 - 2H_2(\qber)\right], \label{eq:bb84}\\
\qber &= \frac{S\,R_\mathrm{det}}{hf\,\eta + S\,R_\mathrm{det}}, \label{eq:qber}
\end{align}
with $Q$ the QBER, $H_2$ the binary entropy function and $hf$ the photon energy, where $h$ is Planck's constant and $f$ the optical frequency of the quantum channel.  The noise power spectral density $S = P_\mathrm{int}/B_\mathrm{ch}$ is the total noise power $P_\mathrm{int}$ that spontaneous Raman scattering, four-wave mixing and transmitter leakage place in the quantum channel, divided by the bandwidth $B_\mathrm{ch}$ of the WDM channel slot. The collection factor of the receiver, $R_\mathrm{det} = B_\mathrm{det}\tau_\mathrm{det}$, is the product of its optical bandwidth and detection gate. We use $R_\mathrm{det} = 1$, the single-temporal-mode limit adopted in~\cite{Zischler2026}, for every experiment, so that IE refers all systems to one reference receiver. Real receivers collect more modes, so IE would change with the actual receiver; NNKR does not depend on this choice. Equation~(\ref{eq:qber}) counts every noise detection as an error. This lowers $R_\mathrm{limit}$ and so gives an upper estimate of IE. The same reference provides the limit for continuous-variable QKD (CV-QKD) with Gaussian-modulated coherent states (GMCS)~\cite{Zischler2026}; multicore fibres are evaluated per core with the single-mode noise model, which neglects the inter-core Raman noise treated in~\cite{Zischler2026_SDM}; in hollow-core fibre the overlap of light with silica is small and the non-linear noise is negligible~\cite{Alia2022}, so $R_\mathrm{limit}$ reduces to its pure-loss value. Physically, $R_\mathrm{limit}$ is the key that perfect equipment would extract from the same channel; the noise penalty of the channel sits in $R_\mathrm{limit}$, and IE is the fraction of that key the real hardware delivers.

\subsection{Relation Between the Two Metrics}
Equations~(\ref{eq:nnkr})--(\ref{eq:bb84}) share $f_\mathrm{rep}$ and $\eta$, so for BB84
\begin{equation}
\frac{\ie}{\nnkr} = \frac{2}{1 - 2H_2(\qber_\mathrm{model})},
\label{eq:ratio}
\end{equation}
where $\qber_\mathrm{model}$ is the noise-induced error rate of~(\ref{eq:qber}). The ratio equals 2 on a noise-free channel and grows with the modelled noise, so it measures the noise penalty directly. A value close to 2 shows that NNKR captures the comparison on its own, and a larger value shows how strongly the noise model shapes it. The measured SKR includes the finite-size penalty, the key reduction that statistical fluctuations impose on a data block of finite size~\cite{Pirandola2020_Advances}, which both metrics include. Table~\ref{tab:reporting} lists the parameters that a publication needs to report for both metrics.

\begin{table}[t]
\centering
\caption{Minimum Reporting Set for NNKR and IE}
\label{tab:reporting}
\footnotesize
\begin{tabular}{@{}lll@{}}
\toprule
\textbf{Parameter} & \textbf{Symbol} & \textbf{Unit} \\
\midrule
\multicolumn{3}{@{}l}{\textit{NNKR}} \\
Secret key rate & $\skr$ & b/s \\
Clock rate & $f_\mathrm{rep}$ & Hz \\
Fibre attenuation; length & $\alpha$; $L$ & dB/km; km \\
\midrule
\multicolumn{3}{@{}l}{\textit{IE (in addition)}} \\
Classical launch power; channel count & $P_\mathrm{WDM}$; $N_\mathrm{ch}$ & dBm; -- \\
Quantum--classical separation & $\Delta\lambda$ & nm \\
Receiver bandwidth; detection gate & $B_\mathrm{det}$; $\tau_\mathrm{det}$ & GHz; ns \\
Propagation direction & -- & co/counter \\
Operating QBER & $\qber$ & \% \\
\midrule
\multicolumn{3}{@{}l}{\textit{Recommended for both}} \\
Component loss beyond $\alpha L$ & -- & dB \\
Key block size & $N$ & bits \\
\bottomrule
\end{tabular}
\end{table}

\section{Mathematical Analysis of the Metrics}
\label{sec:math}

This section derives the properties that make NNKR and IE hardware metrics: an explicit expression for the hardware constant, the uniqueness of the transmittance normalisation, closed-form limits of the loss range over which NNKR is constant, the sensitivity of IE to the noise model, and the propagation of measurement uncertainty.

\subsection{Key-Rate Model and the Hardware Constant}
\label{sec:model}
We describe a decoy-state BB84 link in the standard asymptotic form with an unlimited number of decoy intensities~\cite{Pirandola2020_Advances}. Let $x \equiv \eta$ be the channel transmittance, $\eta_r = T_\mathrm{rx}\eta_\mathrm{det}$ the receiver efficiency, $p_n$ the probability of a noise detection per pulse (dark counts, afterpulses and Raman photons), $e_d$ the intrinsic error probability of a signal detection (set by optical alignment and interferometer visibility), and $f_\mathrm{EC} \geq 1$ the error-correction efficiency factor, equal to 1 at the Shannon limit. For $\mu x \eta_r \ll 1$, the detection probability per signal pulse is $Q_\mu = \mu x \eta_r + p_n$, the single-photon contribution is $Q_1 = \mu e^{-\mu}(x\eta_r + p_n)$, and the error rates of all detections and of single-photon detections are
\begin{equation}
E = \frac{e_d\,\mu x \eta_r + p_n/2}{Q_\mu}, \qquad e_1 = \frac{e_d\,x \eta_r + p_n/2}{x\eta_r + p_n},
\label{eq:errors}
\end{equation}
because a noise detection is in error with probability 1/2. The secret key per pulse is $K = \eta_\mathrm{sift}\,[\,Q_1(1-H_2(e_1)) - f_\mathrm{EC}\,Q_\mu H_2(E)\,]$~\cite{Pirandola2020_Advances}. Introducing the detection signal-to-noise ratio $\rho = \mu x\eta_r/p_n$, the ratio of signal to noise detections, and dividing $K$ by $x$ gives
\begin{equation}
\nnkr = \eta_\mathrm{sift}\,\mu\,\eta_r\,\Phi(\rho),
\label{eq:nnkr_model}
\end{equation}
\begin{equation}
\Phi(\rho) = e^{-\mu}\Big(1+\frac{\mu}{\rho}\Big)\big[1-H_2(e_1)\big] - f_\mathrm{EC}\Big(1+\frac{1}{\rho}\Big)H_2(E),
\label{eq:phi}
\end{equation}
where $\Phi$ is the secure fraction per signal detection, $e_1 = (e_d + \mu/2\rho)/(1+\mu/\rho)$ and $E = (e_d + 1/2\rho)/(1+1/\rho)$.

\textit{Proposition~1 (hardware constant):} as $\rho \to \infty$,
\begin{align}
\nnkr &\to N_0 = \eta_\mathrm{sift}\,\mu\,\eta_r\,\kappa_0, \nonumber\\
\kappa_0 &= e^{-\mu}\big[1-H_2(e_d)\big] - f_\mathrm{EC}H_2(e_d),
\label{eq:N0}
\end{align}
which contains source, receiver, detector and key-processing parameters and is independent of the transmittance. Equation~(\ref{eq:N0}) is the explicit form of the budget (\ref{eq:budget}) with $\kappa = \kappa_0$. For the published parameters of~\cite{Yuan2018} ($\mu = 0.4$, $e_d = 3.07$\%, $f_\mathrm{EC} = 1.34$), (\ref{eq:N0}) gives $\kappa_0 = 0.273$ against a measured secure fraction of 0.292, an agreement within 0.3~dB. Expanding $\Phi$ to first order in $1/\rho$ gives $\nnkr \approx N_0(1 - c_n/\rho)$ with
\begin{multline}
c_n = \frac{1}{\kappa_0}\Big\{f_\mathrm{EC}\big[H_2 + (\tfrac12 - e_d)H_2'\big] \\
- \mu e^{-\mu}\big[1 - H_2 - (\tfrac12 - e_d)H_2'\big]\Big\},
\label{eq:cn}
\end{multline}
where $H_2$ and $H_2' = \log_2[(1-e_d)/e_d]$ are evaluated at $e_d$; for $\mu = 0.5$, $e_d = 2$\% and $f_\mathrm{EC} = 1.16$, $\kappa_0 = 0.357$ and $c_n = 10.8$, so the noise correction stays below 1~dB once signal detections outnumber noise detections by about 50 to 1.

Detectors with a dead time $t_d$ recover non-paralysably; with $N_d$ detectors sharing the detections, the detected fraction, or saturation factor, is
\begin{equation}
F_\mathrm{sat}(x) = \frac{1}{1 + x/x_s}, \qquad x_s = \frac{N_d}{f_\mathrm{rep}\,\mu\,\eta_r\,t_d},
\label{eq:sat}
\end{equation}
where $x_s$ is the transmittance at which dead time halves the detection rate; a key-processing throughput ceiling produces the same functional form. Combining (\ref{eq:nnkr_model}) and (\ref{eq:sat}),
\begin{align}
\nnkr(x) &= N_0\,G(x/x_n)\,F_\mathrm{sat}(x), \nonumber\\
G &= \frac{\Phi}{\kappa_0}, \qquad x_n = \frac{p_n}{\mu\,\eta_r},
\label{eq:full}
\end{align}
where $G$ is the noise factor, which equals 1 when noise detections vanish, and $x_n$ is the transmittance at which noise and signal detections are equal, so that $\rho = x/x_n$.

\subsection{Uniqueness of the Transmittance Normalisation}
\label{sec:unique}
\textit{Proposition~2 (uniqueness):} consider any metric $M = K/\psi(\eta)$, where $K = \skr/f_\mathrm{rep}$ is the key per pulse and $\psi$ a normalisation function. In the loss-limited regime, $K = N_0\,\eta$ for every system. $M$ is then independent of $\eta$ for every $N_0$ if and only if $\psi(\eta) = a\,\eta$ with a constant $a$, and NNKR is the choice $a = 1$. \textit{Proof:} $M = N_0\,\eta/\psi(\eta)$ is constant in $\eta$ exactly when $\eta/\psi(\eta)$ is constant. $\blacksquare$ The same argument applies to protocols whose key scales as $\eta^\beta$: their invariant normalisation is $K/\eta^\beta$, with $\beta = 1/2$ for twin-field QKD, so the framework carries over with $\eta$ replaced by $\eta^\beta$.

\begin{figure}[t]
\centering
\includegraphics[width=\columnwidth]{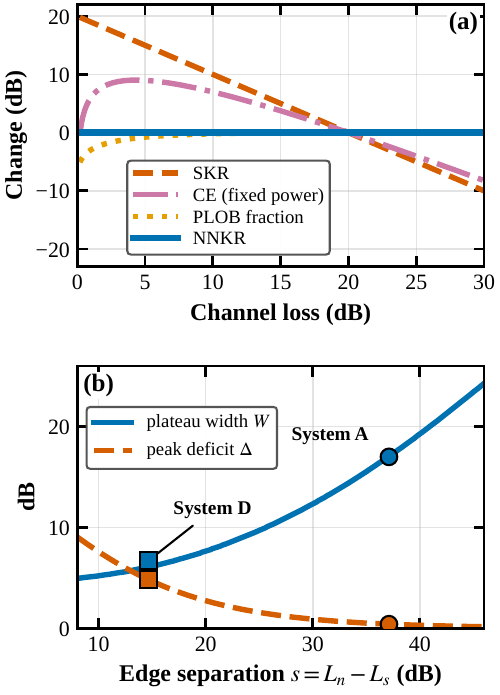}
\caption{Analytic properties. (a) Change of each metric with channel loss for fixed loss-limited hardware, relative to its value at 20~dB: SKR falls by 1~dB per dB, CE (fixed classical power) peaks at 4.34~dB of loss, the PLOB fraction falls at low loss by $\varphi(\eta)$ of (\ref{eq:phi_plob}), and NNKR stays constant (Proposition~2). (b) Width $W$ of the 1-dB plateau and deficit $\Delta$ of its maximum below the hardware constant $N_0$, as functions of the edge separation $s$ of (\ref{eq:width}); markers show Systems~A and~D, with measured widths (blue) and model deficits (orange).}
\label{fig:theory}
\end{figure}

The established metrics correspond to other choices of $\psi$, and Proposition~2 quantifies their response to loss for fixed hardware (Fig.~\ref{fig:theory}(a)). For the PLOB fraction, $\psi = -\log_2(1-\eta)$, so
\begin{equation}
\frac{M_\mathrm{PLOB}}{\nnkr} = \varphi(\eta) = \frac{\eta}{-\log_2(1-\eta)},
\label{eq:phi_plob}
\end{equation}
which tends to $\ln 2$ at high loss and falls as the loss falls: relative to its high-loss value, the PLOB fraction of fixed hardware is 2.0~dB lower at 2~dB of loss, 0.8~dB lower at 5~dB and 0.2~dB lower at 10~dB. For CE at fixed classical power, writing the channel loss as $A = \alpha L$ in decibels,
\begin{align}
\ce &= f_\mathrm{rep}\,N_0\,P_\mathrm{WDM}\,\frac{A}{\alpha}\,10^{-A/10}, \nonumber\\
\frac{d\ln\ce}{dA} &= \frac{1}{A} - \frac{\ln 10}{10},
\label{eq:ce_A}
\end{align}
which vanishes at $A^\ast = 10/\ln 10 = 4.34$~dB, that is $L^\ast = 4.34/\alpha$; CE of fixed hardware therefore rises by a factor of 8.4 (9.2~dB) between 0.2 and 4.3~dB of loss and falls beyond. The SKR itself falls by 1~dB per dB of loss. Of these quantities, NNKR alone stays flat across the full range.

\subsection{Edges and Width of the NNKR Plateau}
\label{sec:edges}
\textit{Proposition~3 (plateau edges):} for a system described by (\ref{eq:full}), let $L_s = -10\log_{10}x_s$ and $L_n = -10\log_{10}x_n$ be the saturation and noise losses, and let $\rho_1$ solve $G(\rho_1) = 10^{-0.1}$. When the two edges are well separated, NNKR stays within 1~dB of $N_0$ from $L_\mathrm{lo} = L_s + 10\log_{10}[1/(10^{0.1}-1)] = L_s + 5.87$~dB to $L_\mathrm{hi} = L_n - 10\log_{10}\rho_1$, a width of
\begin{align}
W_0 &= s - 10\log_{10}\rho_1 - 5.87~\mathrm{dB}, \nonumber\\
s &= L_n - L_s = 10\log_{10}\frac{N_d}{R_n\,t_d},
\label{eq:width}
\end{align}
where $R_n = f_\mathrm{rep}\,p_n$ is the noise count rate. \textit{Proof:} $F_\mathrm{sat}(x) = 10^{-0.1}$ at $x = (10^{0.1}-1)\,x_s$ and $G = 10^{-0.1}$ at $\rho = \rho_1$, that is $x = \rho_1 x_n$; the expression for $s$ follows from (\ref{eq:sat}) and $x_n = p_n/(\mu\eta_r)$. $\blacksquare$

The edge separation $s$ depends on the noise count rate per detector multiplied by the dead time, and on nothing else: the loss range over which NNKR is constant is a property of the detectors. The constant $\rho_1$ is 33--96 (15.2--19.8~dB) for $\mu = 0.4$--0.6, $e_d = 0.5$--3\% and $f_\mathrm{EC} = 1.05$--1.3; with $\rho_1 = 50$, $W_0 = s - 22.9$~dB. When the edges approach each other, the maximum of NNKR falls below $N_0$ by a deficit $\Delta$, and the 1-dB plateau measured relative to that maximum has width $W$; Fig.~\ref{fig:theory}(b) gives both as functions of $s$ from (\ref{eq:full}). Two regimes follow. For $s \geq 35$~dB, that is $R_n t_d/N_d \leq 3\times10^{-4}$, $\Delta \leq 0.5$~dB and the plateau level equals the hardware constant within 0.5~dB. For smaller $s$, the plateau level is a lower bound on $N_0$, below it by $\Delta$, which (\ref{eq:full}) supplies.

\subsection{Noise Sensitivity of the Implementation Efficiency}
\label{sec:elastic}
From (\ref{eq:ratio}) and (\ref{eq:qber}), with $\partial \qber/\partial \ln S = \qber(1-\qber)$, the elasticity of IE with respect to the modelled noise power spectral density $S$ is
\begin{equation}
\varepsilon_S = \frac{\partial \ln \ie}{\partial \ln S} = \frac{2\,\qber(1-\qber)\log_2[(1-\qber)/\qber]}{1-2H_2(\qber)},
\label{eq:elastic}
\end{equation}
with $\qber = \qber_\mathrm{model}$. $\varepsilon_S$ equals 0.02, 0.05, 0.16 and 0.48 at $\qber_\mathrm{model} = 0.1$\%, 0.3\%, 1\% and 3\%, so a relative error $\delta$ in the modelled noise changes IE by about $\varepsilon_S\,\delta$. For the published experiments, where $\qber_\mathrm{model} \leq 0.6$\%, a factor-of-two change in the modelled noise changes IE by at most 10\%; for example, for Gavignet'24, doubling the modelled noise raises IE by 9.7\% and halving it lowers IE by 4.9\%, because $R_\mathrm{limit}$ falls as the noise grows. Above $\qber_\mathrm{model} \approx 1$\%, the accuracy of IE follows the accuracy of the noise model with weight $\varepsilon_S$, and calibrated noise coefficients become part of the measurement.

\subsection{Error Propagation and Statistical Consistency}
\label{sec:errors}
Since $\nnkr = \skr/(f_\mathrm{rep}\,\eta)$ is a product of independent factors, its standard uncertainty in decibels is
\begin{equation}
\sigma_\mathrm{NNKR}^2 = \sigma_A^2 + \Big(\frac{10}{\ln 10}\Big)^2\Big[\Big(\frac{\sigma_R}{\skr}\Big)^2 + \Big(\frac{\sigma_f}{f_\mathrm{rep}}\Big)^2\Big],
\label{eq:sigma}
\end{equation}
where $\sigma_A$ is the uncertainty of the channel loss in decibels and $\sigma_R$ and $\sigma_f$ are the standard uncertainties of the SKR and the clock rate. The clock term is negligible for crystal-referenced transmitters, so NNKR inherits the loss uncertainty one for one and the relative SKR uncertainty with a weight of 4.34~dB; for the 13.72~Mb/s record, the reported $\pm5.4$\% SKR fluctuation contributes $\pm0.23$~dB~\cite{Yuan2018}. Under (\ref{eq:sigma}), the agreement of $J$ independent determinations $N_j$ of the same hardware, with mean $\bar N$, is tested by $\chi^2 = \sum_j (N_j - \bar N)^2/\sigma^2$ with $J-1$ degrees of freedom. The flatness of NNKR is quantified by its loss sensitivity $b$, the slope of NNKR in decibels against channel loss in decibels: ideal invariance gives $b = 0$, and the SKR has $b = -1$.

\section{Analysis Method and Substantiation}
\label{sec:method}

The analysis draws on four data sets. The literature set comprises ten published coexistence demonstrations that report SKR, fibre length and classical launch power, selected to span four protocols (decoy-state BB84, coherent one-way (COW), CV-QKD and MDI-QKD), four fibre types (dispersion-shifted, standard single-mode, multicore and hollow-core), the O band (1260--1360~nm) and C band (1530--1565~nm), and classical capacities from 10~Gb/s to 18~Tb/s (Table~\ref{tab:experiments}); multicore results are evaluated per core. The state-of-the-art set comprises four high-rate laboratory systems whose clock rate, channel loss and SKR are stated in their publications, including the multi-distance measurements of two of them~\cite{Yuan2018,Li2023,Grunenfelder2023,FanYuan2026}. The testbed set comprises 42{,}115 SKR values logged by four QKD systems on the TCD/CONNECT testbed during automated attenuation sweeps over ITU-T G.652.D standard single-mode fibre, with dense wavelength-division multiplexing (DWDM) multiplexers, circulators and band filters representative of a production link. System~A is a Toshiba production unit and System~B a Toshiba research unit with external fibre patching, both GHz-clocked decoy-state BB84 with indium gallium arsenide (InGaAs) avalanche photodiodes; System~C is an ID~Quantique (IDQ) Clavis3 running the COW protocol at 625~MHz; System~D is an ID~Quantique ClavisXGR running GHz-clocked decoy-state BB84. The coexistence set comprises a Toshiba system connected through a reconfigurable optical add-drop multiplexer (ROADM), measured over 0 and 50~km of fibre with a variable optical attenuator (VOA) sweep at baseline and with 12~dBm (16~mW) of classical injection, from the Irish national quantum communication infrastructure project (IrelandQCI; Data Availability).

For every entry, CE follows from $\ce = \skr\,P_\mathrm{WDM}\,L$~\cite{Gavignet2024_JLT}, NNKR from (\ref{eq:nnkr}) and, for the BB84 and CV-QKD entries, IE from (\ref{eq:ie})--(\ref{eq:qber}); on the testbed, $\eta$ is the set channel attenuation. To identify the loss-limited range of each system with an objective criterion, we define its \emph{1-dB plateau}, by analogy with the 1-dB bandwidth of an optical component, as the contiguous loss range that contains the NNKR maximum and over which NNKR, median-filtered over 1~dB, stays within 1~dB of that maximum. The plateau level is the hardware value of the system, its width is the loss range over which the system is channel-limited, and its coefficient of variation (CoV), the standard deviation divided by the mean, quantifies flatness. The same criterion applies to published multi-distance data.

Four checks substantiate the computation. First, CE recomputed from the reported SKR, launch power and length reproduces all ten entries of Table~\ref{tab:experiments} to within 2\%. Second, the IE column satisfies (\ref{eq:ratio}) entry by entry: IE/NNKR equals 2.00--2.03 for the five low-noise BB84 experiments, where the modelled noise error rate is below 0.1\%, and 2.23 for Gavignet'24, the BB84 experiment with the highest classical power; for Hugues-Salas'20, Table~\ref{tab:experiments} lists the noise-free bound. Third, the budget of the 13.7~Mb/s record closes within 0.3~dB (Section~\ref{sec:sota}), and an independent check at the sifted-key level gives a predicted $\mu T_\mathrm{rx}\eta_\mathrm{det}\eta_\mathrm{sift} = 0.070$ against a measured 0.076 sifted bits per surviving pulse~\cite{Yuan2018}, a 0.3~dB difference consistent with the 4.4\% afterpulse probability and 450~kHz dark-count rate reported for those detectors. Fourth, the testbed value of System~A agrees with three separately published deployments of the same platform within 0.7~dB (Section~\ref{sec:lab}).

The model of Section~\ref{sec:math} is fitted to the testbed curves by robust least squares in decibels, with $N_0$, $L_n$ and $L_s$ as free parameters and $\mu = 0.5$, $e_d = 2$\% and $f_\mathrm{EC} = 1.16$ fixed; within the parameter ranges of Section~\ref{sec:edges}, these fixed values shift the fitted edges by less than 1~dB. Uncertainties follow (\ref{eq:sigma}), loss sensitivities are least-squares slopes over each plateau, and cross-laboratory agreement is tested with the $\chi^2$ statistic of Section~\ref{sec:errors}. For the coexistence set, each metric is compared with its baseline value at the same fibre length and VOA setting. The clock rate, the transmittance and the component losses then cancel. The per-point uncertainty is the 6.9\% CoV of System~A, which uses the same platform. With (\ref{eq:sigma}), this gives 0.42~dB for each change and $0.42/\sqrt{n}$~dB for the mean of $n$ changes. The figure data and the scripts that produce every figure and statistic in this paper are available from the corresponding author on request.

\section{Numerical Results}
\label{sec:results}

\subsection{Ten Coexistence Experiments}
\label{sec:lit}

\begin{figure}[t]
\centering
\includegraphics[width=0.92\columnwidth]{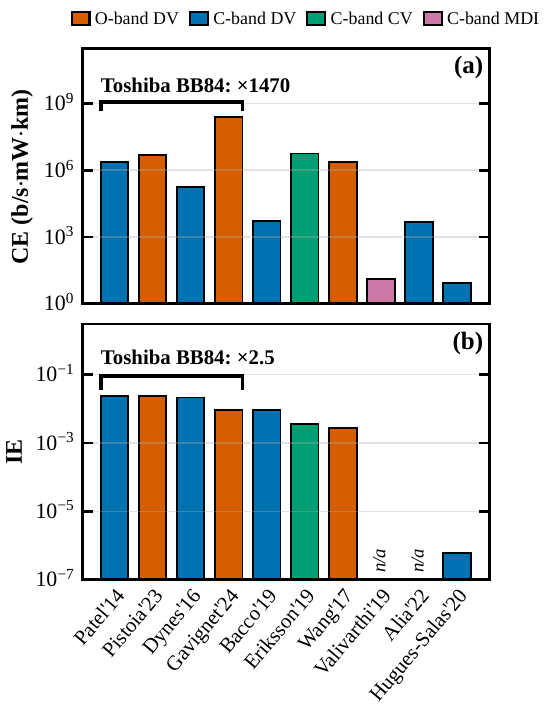}
\caption{(a) Co-propagation efficiency and (b) implementation efficiency of the ten experiments in Table~\ref{tab:experiments}, in the same order. The bracket marks the four Toshiba BB84 deployments: $\times$1470 under CE, which scales with classical power and length, and $\times$2.5 under IE. IE is not shown for the COW and MDI-QKD experiments, for which no coexistence-aware key-rate limit is available. DV: discrete-variable; CV: continuous-variable; MDI: measurement-device-independent.}
\label{fig:ce_ie}
\end{figure}

Fig.~\ref{fig:ce_ie} plots CE for the ten experiments of Table~\ref{tab:experiments} and IE for the eight with an available limit, in the same order. CE spans $3\times10^{7}$ and varies by a factor of 1470 across the four Toshiba BB84 deployments. IE places the three earlier ones, Patel'14, Dynes'16 and Pistoia'23, within a factor of 1.09 ($\ie = 0.021$--$0.023$) across 50--70~km of fibre, with classical traffic from 10~Gb/s to 2.4~Tb/s and both O- and C-band quantum channels, and places Gavignet'24, which carried the highest classical power, below them; across all four, the IE spread is a factor of 2.5. The ranking changes accordingly: Patel'14 moves from fifth under CE to first under IE, and Gavignet'24 moves from first to fourth. Physically, IE follows the per-pulse efficiency of the equipment, which is set by its detectors, receiver optics, source settings and key processing; these properties travel with the hardware, so one platform gives one IE value in different fibres, bands and traffic loads, provided the classical power does not require a different operating configuration.

The three earlier deployments kept the classical noise at the receiver low for their band. Patel'14 and Dynes'16 placed the quantum channel in the C band with $-14$ and $-25.5$~dBm of classical power and combined spectral filtering with the temporal filtering of GHz-gated detectors~\cite{Patel2014,Dynes2016}; Pistoia'23 placed it in the O band, where adding nine channels to reach 2.4~Tb/s moved the QBER from 3.95\% to 4.04\%~\cite{Pistoia2023}. Their IE values agree within a factor of 1.09 (0.4~dB), a coefficient of variation of 4.9\%, below the 6.9\% that System~A shows along its own plateau (Section~\ref{sec:lab}). Three deployments, two bands and classical capacities from 10~Gb/s to 2.4~Tb/s thus give one IE value, and the classical traffic leaves no measurable penalty in it. Gavignet'24 used the same platform to carry 6~Tb/s at 16.8~dBm, the highest classical power of the ten experiments, and its CE exceeds theirs by a factor of 54 to 1470. Its IE lies 3.8~dB below theirs. The two metrics describe one trade-off: the system was configured and operated to carry very high classical power alongside the quantum channel, and the cost appears as a lower IE. The authors report the same trade-off directly: the SKR falls from 169 to 106~kb/s when the classical channels are added~\cite{Gavignet2024_JLT}. IE shows this difference without needing the details of the configuration, and it leaves the three earlier deployments unpenalised. Across the BB84 entries, the modelled Raman noise stays small: IE/NNKR equals 2.00--2.03 for the five low-noise experiments and 2.23 for Gavignet'24, a modelled noise error rate of 0.6\%, so the published experiments operate in the signal-dominated regime, where IE and $2\cdot$NNKR agree closely.

\subsection{Comparison With State-of-the-Art Systems}
\label{sec:sota}

\begin{figure}[t]
\centering
\includegraphics[width=0.92\columnwidth]{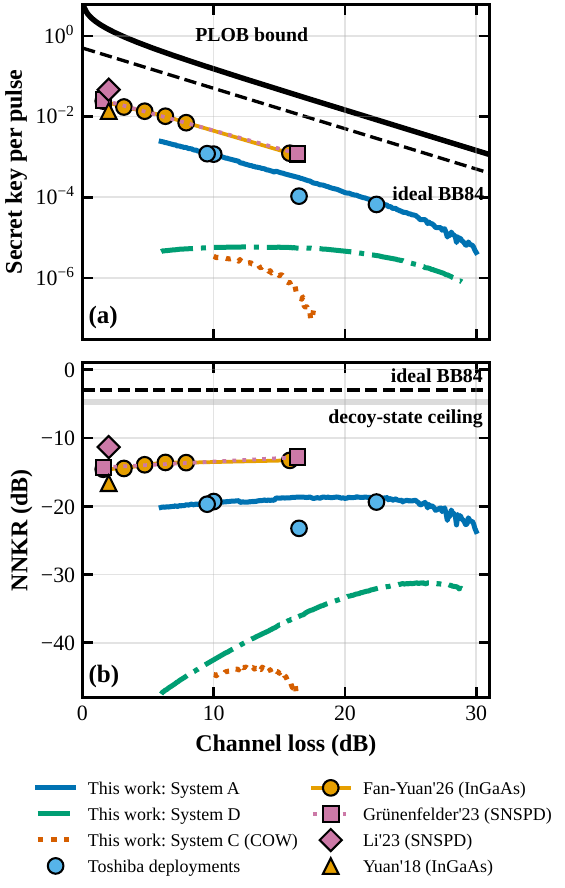}
\caption{Comparison with state-of-the-art systems. (a) Conventional representation: secret key per pulse against channel loss, with the PLOB bound and ideal BB84. (b) The same data as NNKR in decibels: the exponential channel loss is divided out, each system becomes a near-horizontal line at its hardware value, and the distance to the ideal lines is its penalty budget. Records: Li'23~\cite{Li2023}, Gr\"unenfelder'23~\cite{Grunenfelder2023}, Fan-Yuan'26~\cite{FanYuan2026}, Yuan'18~\cite{Yuan2018}; Toshiba deployments from Table~\ref{tab:experiments}. PLOB: repeaterless capacity bound; SNSPD: superconducting nanowire single-photon detector; InGaAs: indium gallium arsenide avalanche photodiode; COW: coherent one-way.}
\label{fig:sota}
\end{figure}

Fig.~\ref{fig:sota} compares the testbed systems and the published deployments with four high-rate laboratory records, two built with superconducting nanowire single-photon detectors (SNSPDs) and two with InGaAs avalanche photodiodes, in two representations. Panel~(a) uses the conventional representation, secret key per pulse against channel loss with the PLOB bound and ideal BB84 as references: every system falls exponentially with loss, and differences in hardware appear as small vertical offsets on a logarithmic scale that spans seven decades. Panel~(b) plots the same data as NNKR in decibels, which divides out the exponential channel loss: each system becomes a nearly horizontal line whose height is its hardware value, and the vertical distance to the ideal lines is its penalty budget. Table~\ref{tab:sota} lists the inputs.

Three results emerge from this comparison. First, the record systems themselves show the plateau behaviour of Section~\ref{sec:nnkr}. The six distances reported by Fan-Yuan \textit{et al.}~\cite{FanYuan2026} give a 1-dB plateau at $-13.6$~dB from 4.7 to 15.8~dB of loss, and the 10-km record point lies 1.0~dB lower, at $-14.6$~dB, where the authors identify detector afterpulsing as the limit. Gr\"unenfelder \textit{et al.}~\cite{Grunenfelder2023} rise by 1.5~dB from their 10-km point to their 102-km point, and their reported detector efficiency, 64\% at the 320 million counts per second (Mcount/s) of the short link against 81\% at the 15~Mcount/s of the long link, accounts for 1.0~dB of this rise. The 10-km benchmark at which SKR records are quoted therefore places high-rate systems in their detector-saturation regime, and NNKR read on the plateau gives their hardware value, 1.0--1.5~dB above the record point.

\begin{figure}[t]
\centering
\includegraphics[width=\columnwidth]{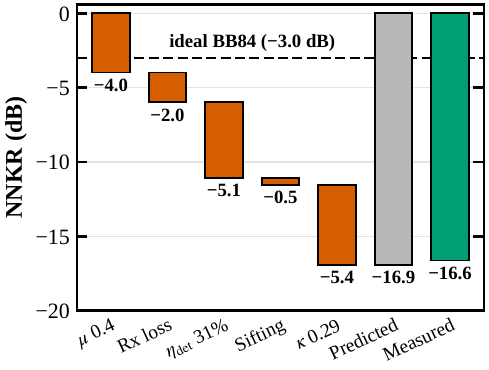}
\caption{NNKR as a penalty budget for the 13.7~Mb/s system of~\cite{Yuan2018}: the five hardware factors of (\ref{eq:budget}), taken from the published parameters, predict $-16.9$~dB against a measured $-16.6$~dB. $\mu$: mean photon number; Rx: receiver; $\eta_\mathrm{det}$: detector efficiency; $\kappa$: secure fraction.}
\label{fig:budget}
\end{figure}

Second, the scale decomposes into identifiable hardware contributions. Fig.~\ref{fig:budget} evaluates (\ref{eq:budget}) for the 13.7~Mb/s system of Yuan \textit{et al.}~\cite{Yuan2018} from its published parameters: a mean photon number of 0.4 ($-4.0$~dB), a receiver interferometer loss of 2~dB, a detector efficiency of 31\% ($-5.1$~dB), a sifting efficiency of 0.90 ($-0.5$~dB) and a secure fraction of 0.29 after privacy amplification and discarded error-correction blocks ($-5.4$~dB) sum to $-16.9$~dB, against a measured $-16.6$~dB. Against this reference, Fan-Yuan \textit{et al.} use 40\%-efficient InGaAs detectors and a 0.3~dB receiver~\cite{FanYuan2026}, which predict a 2.8~dB advantage and account for the observed 2.0~dB at 10~km within 0.8~dB; Gr\"unenfelder \textit{et al.} combine 64\% detector efficiency with a secure fraction of 0.40, together worth 4.5~dB, and the observed 2.3~dB places the remaining 2.2~dB in the mean photon number, sifting and receiver terms, consistent with the timing-jitter loss of 100-ps detection windows at 2.5~GHz that the authors identify.

Third, production and record systems fall on one scale. The production deployments and System~A sit at $-19.0$ to $-19.7$~dB, 2.4--3.1~dB below the 13.7~Mb/s research system of the same vendor~\cite{Yuan2018} and 5.4~dB below the InGaAs plateau of Fan-Yuan \textit{et al.}; by (\ref{eq:budget}), the 5.4~dB corresponds to a 3.5-fold lower product of the five hardware factors. The superconducting-detector record of Li \textit{et al.} reaches $-11.3$~dB at its 10-km point, and between it and the decoy-state ceiling lie 6--7~dB, the remaining headroom in detector efficiency, timing jitter, error-correction efficiency and finite-size effects. The same scale resolves the lowest tier of Table~\ref{tab:experiments}: Alia'22, Valivarthi'19 and Hugues-Salas'20 lie 23--46~dB below the production cluster. MDI-QKD carries an intrinsic penalty of about 8~dB relative to BB84, because every key bit requires two photons from independent sources to meet at a Bell-state measurement, the joint two-photon measurement at the central node, which succeeds with 50\% probability, giving $(\mu e^{-\mu})^2/2 \approx 0.045$ against $\mu e^{-\mu} \approx 0.30$ for $\mu = 0.5$; COW encodes each bit in a pair of time slots and taps part of the light for coherence monitoring; testbed System~C, the same COW platform (IDQ Clavis3) as Alia'22, sits at $-43.9$~dB, and the 7.8~dB between them is comparable to the 7.9~dB of multiplexer and filter loss that Alia \textit{et al.} report beyond the fibre (10.5~dB end to end against 2.6~dB of fibre)~\cite{Alia2022}, which NNKR counts as hardware; and Hugues-Salas'20 operated at a QBER of 6.3\%, where the BB84 key fraction $1-2H_2(\qber)$ falls to 0.32, a 4.9~dB reduction of $\kappa$, in a seven-core fibre carrying 11.2~Tb/s.

\begin{table}[t]
\centering
\caption{NNKR of State-of-the-Art Systems, Production Deployments and This Work}
\label{tab:sota}
\footnotesize\setlength{\tabcolsep}{2.6pt}
\begin{tabular}{@{}lcccccc@{}}
\toprule
\textbf{System} & \textbf{Detector} & \textbf{$\eta_\mathrm{det}$} & \textbf{$f_\mathrm{rep}$} & \textbf{Loss} & \textbf{SKR} & \textbf{NNKR} \\
 & & (\%) & (GHz) & (dB) & (Mb/s) & (dB) \\
\midrule
Li'23~\cite{Li2023} & SNSPD & -- & 2.5 & 2.0$^a$ & 115.8 & $-11.3$ \\
Gr\"unenfelder'23~\cite{Grunenfelder2023} & SNSPD & 64 & 2.5 & 1.6$^b$ & 64 & $-14.3$ \\
\quad at 102.4~km & SNSPD & 81 & 2.5 & 16.4$^b$ & 3.0 & $-12.8$ \\
Fan-Yuan'26~\cite{FanYuan2026} & InGaAs & 40 & 2.5 & 1.58 & 60.33 & $-14.6$ \\
\quad 1-dB plateau & InGaAs & 40 & 2.5 & 4.7--15.8 & 33.9--3.1 & $-13.6$ \\
Yuan'18~\cite{Yuan2018} & InGaAs & 31 & 1.0 & 2.0 & 13.72 & $-16.6$ \\
\midrule
Patel'14~\cite{Patel2014} & InGaAs & -- & 1.0 & 10.0 & 1.17 & $-19.3$ \\
Dynes'16~\cite{Dynes2016} & InGaAs & -- & 1.0 & 9.5 & 1.2 & $-19.7$ \\
Pistoia'23~\cite{Pistoia2023} & InGaAs & -- & 1.0 & 22.4 & 0.066 & $-19.4$ \\
Gavignet'24~\cite{Gavignet2024_JLT} & InGaAs & -- & 1.0 & 16.0 & 0.106 & $-23.7$ \\
This work, Sys.~A & InGaAs & -- & 1.0 & 8.9--25.9 & 1.4--0.03 & $-19.0$ \\
\bottomrule
\multicolumn{7}{@{}p{\columnwidth}@{}}{\footnotesize SNSPD: superconducting nanowire single-photon detector; InGaAs: indium gallium arsenide avalanche photodiode; $\eta_\mathrm{det}$: detector efficiency, as stated at the operating point; $f_\mathrm{rep}$: clock rate. Loss for the production deployments follows from their reported fibre length and attenuation. $^a$10~km of standard fibre at 0.2~dB/km, as in~\cite{FanYuan2026}. $^b$Ultra-low-loss fibre at 0.16~dB/km, the attenuation of the ultra-low-loss fibre in~\cite{FanYuan2026}.}
\end{tabular}
\end{table}

\subsection{Comparison With State-of-the-Art Figures of Merit}
\label{sec:compare}

Table~\ref{tab:metric_comparison} compares NNKR and IE quantitatively with the established figures of merit by applying each one to four sets of measurements: one platform across four deployments, the three earlier deployments of that platform, System~A across its plateau, and the six distances of Fan-Yuan \textit{et al.} CE and the PLOB fraction were designed for other purposes, coexistence capacity and distance to capacity; Table~\ref{tab:metric_comparison} shows only how much each responds to operating conditions when the hardware is nominally fixed. For fixed hardware a hardware metric should stay constant, so the spread, the ratio of the largest to the smallest value, measures how much of each metric reflects operating conditions. SKR varies by factors of 18 to 49, because it includes the channel loss.

CE varies by a factor of 1470 across the Toshiba deployments because it multiplies the SKR by classical launch power and length~\cite{Gavignet2024_JLT}, which makes it a measure of coexistence capacity. Three operating conditions set its value. The wavelength band dominates: an O-band quantum channel sits about 35~THz above C-band classical channels, on the anti-Stokes side of the Raman spectrum (frequencies above the classical channels), where the Raman noise is about 4000 times (36~dB) weaker than at 1550~nm~\cite{Eraerds2010}, and across Table~\ref{tab:experiments} CE spans $3\times10^{7}$, from 8.2 for Hugues-Salas'20 to $2.5\times10^{8}$ for Gavignet'24, both commercial systems. The launch power enters in linear units, so 16.8~dBm (47.9~mW) and $-25.5$~dBm (0.0028~mW) differ by a factor of 17{,}000. Distance enters through (\ref{eq:ce_A}): for a loss-limited link CE peaks at $L^\ast = 4.34/\alpha$, that is 21.7~km at 0.2~dB/km, where the fractional photon loss per kilometre equals $1/L$, and the same hardware gives an eightfold higher CE at 22~km than at 1~km.

The PLOB fraction, the key per pulse divided by the repeaterless capacity $-\log_2(1-\eta)$~\cite{Pirandola2017_PLOB}, divides out both clock rate and loss and agrees with NNKR at high loss, where the capacity approaches $1.44\,\eta$: its spreads of 2.75 and 1.35 for the Toshiba deployments and System~A match those of NNKR. At low loss the capacity grows faster than $\eta$, and (\ref{eq:phi_plob}) gives the resulting bias: across the Fan-Yuan distances the PLOB fraction varies by a factor of 2.27, against 1.35 for NNKR, and at the 2-dB channel of the 13.7~Mb/s record it sits 2.0~dB below NNKR$/1.44$, as $\varphi(\eta)$ predicts. NNKR has no such loss bias; its spread of 1.35 across the Fan-Yuan distances comes mainly from the saturated 10-km point (Section~\ref{sec:sota}). NNKR is therefore the representation that holds hardware constant from metro links at 2~dB to long links at 25~dB. Bebrov's efficiency~\cite{Bebrov2024}, which counts consumed resources, and the quality-of-service parameters of ITU-T Y.3807~\cite{ITUT_Y3807_2022}, which describe service delivery, address protocol and service questions and complement the hardware view. Because $\skr = f_\mathrm{rep}\cdot\eta\cdot\nnkr$, reporting $f_\mathrm{rep}$ alongside NNKR gives the full picture of a system: NNKR describes the receiver, detectors and key processing, $f_\mathrm{rep}$ the transmitter speed, and $\eta$ the channel.

\begin{table*}[t]
\centering
\caption{Comparison With State-of-the-Art Figures of Merit: Spread (Max/Min) of Each Metric for Fixed Hardware}
\label{tab:metric_comparison}
\footnotesize\setlength{\tabcolsep}{4pt}
\begin{tabular}{@{}lccccccl@{}}
\toprule
\textbf{Metric} & \textbf{Loss and clock} & \textbf{Noise} & \textbf{Toshiba, four} & \textbf{Toshiba, three} & \textbf{System A,} & \textbf{Fan-Yuan'26,} & \textbf{Main use} \\
 & \textbf{divided out} & \textbf{modelled} & \textbf{deployments} & \textbf{earlier deployments} & \textbf{8.9--25.9 dB} & \textbf{1.6--15.8 dB} & \\
\midrule
SKR & \nope & \nope & 18 & 18 & 49 & 20 & Link performance \\
CE~\cite{Gavignet2024_JLT} & \nope & \nope & 1470 & 27 & \nope$^a$ & \nope$^a$ & Coexistence capacity \\
PLOB fraction~\cite{Pirandola2017_PLOB} & \yes & \nope & 2.75 & 1.14 & 1.35 & 2.27 & Distance to capacity \\
\textbf{NNKR (this work)} & \yes & \nope & \textbf{2.77} & \textbf{1.09} & \textbf{1.31} & \textbf{1.35} & \textbf{Hardware, four inputs} \\
\textbf{IE (this work)} & \yes & \yes & \textbf{2.49} & \textbf{1.09} & \textbf{1.31}$^b$ & \textbf{1.35}$^b$ & \textbf{Hardware, noise-aware} \\
\bottomrule
\multicolumn{8}{@{}p{\textwidth}@{}}{\footnotesize Spread: largest divided by smallest value of the metric over the measurement set. PLOB fraction: secret key per pulse divided by the PLOB capacity $-\log_2(1-\eta)$. $^a$Standalone measurements, $P_\mathrm{WDM} = 0$. $^b$Standalone, IE $= 2\cdot$NNKR. The four Toshiba deployments include Gavignet'24, whose hardware configuration differs from the three earlier deployments, which were operated at low classical power for their respective bands (Section~\ref{sec:lit}).}
\end{tabular}
\end{table*}

\subsection{Testbed Measurements}
\label{sec:lab}

\begin{figure}[t]
\centering
\includegraphics[width=0.92\columnwidth]{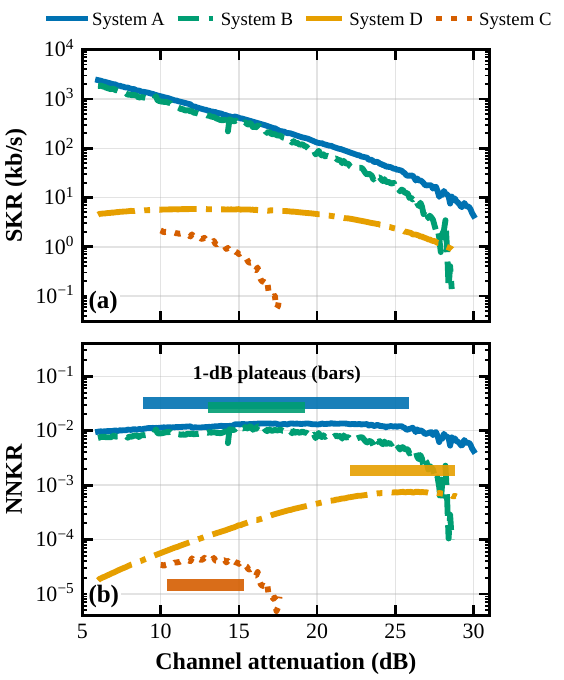}
\caption{Testbed measurements of four QKD systems against channel attenuation: (a) SKR and (b) NNKR. The bars in (b) mark the 1-dB plateau of each system (Table~\ref{tab:testbed}); System~D's NNKR rises while its SKR is rate-limited and becomes flat once it is loss-limited above 22~dB.}
\label{fig:testbed}
\end{figure}

Fig.~\ref{fig:testbed}(a) shows the SKR of the four testbed systems, from 2~Mb/s for System~A at 6~dB to below 1~b/s for System~C at 17~dB, and Fig.~\ref{fig:testbed}(b) shows their NNKR together with the 1-dB plateaus, which Table~\ref{tab:testbed} summarises. Every system shows a plateau. System~A holds NNKR at 0.0125 ($-19.0$~dB) over a 17.0~dB plateau from 8.9 to 25.9~dB with a CoV of 6.9\%. System~B, the research unit of the same vendor, plateaus at $-19.9$~dB over 6.2~dB. System~D delivers 4.7--5.9~kb/s from 6 to 20~dB, nearly independent of loss, so its NNKR rises by a factor of 26 over that range and then settles into a plateau at $-31.5$~dB from 22.1 to 28.8~dB with a CoV of 5.4\%. System~C plateaus at $-43.9$~dB over 4.9~dB before it reaches its range limit near 17.6~dB.

These behaviours follow from the physics of photon detection. A flat NNKR means that each photon reaching the receiver yields the same amount of key: signal photons dominate the detections, the error rate is stable, and error correction costs the same fraction of bits at every loss. Above the upper edge of a plateau, dark counts and afterpulses approach the signal detection rate, the error rate rises and NNKR rolls off toward the reach limit, which for System~A begins at 26~dB. Table~\ref{tab:testbed} quantifies the invariance: over each plateau, the loss sensitivity of NNKR is $+0.017$ to $+0.035$~dB/dB, against $-0.98$ to $-1.01$~dB/dB for the SKR, so NNKR removes 96--98\% of the loss dependence; the record of Fan-Yuan \textit{et al.} gives the same result (95\%) on independent data. The small positive slopes are the residual low-loss saturation factor of (\ref{eq:sat}).

\begin{figure}[t]
\centering
\includegraphics[width=0.92\columnwidth]{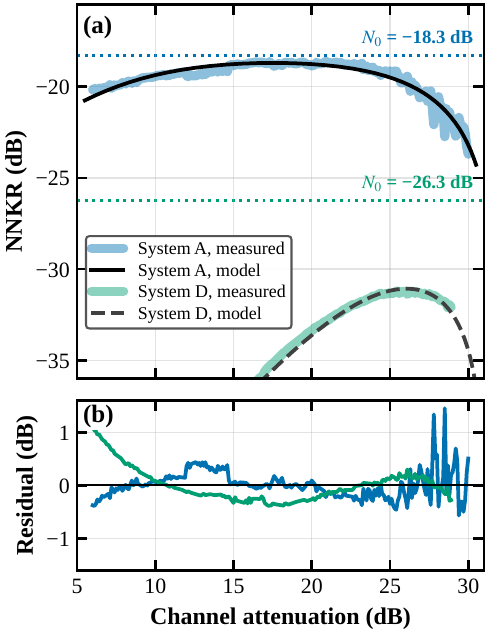}
\caption{Physical model of the testbed curves. (a) Measured NNKR of Systems~A and~D and the fitted model (\ref{eq:full}); dotted lines mark the fitted hardware constants $N_0$. System~A reaches its constant within 0.4~dB; the rate ceiling of System~D keeps its maximum 4.8~dB below its constant. (b) Residuals, with root-mean-square values of 0.28~dB (A) and 0.33~dB (D).}
\label{fig:fit}
\end{figure}

The model (\ref{eq:full}) reproduces the measured curves quantitatively (Fig.~\ref{fig:fit}). For System~A, the fit gives $N_0 = -18.3$~dB, $L_s = 4.3$~dB and $L_n = 41.4$~dB, with a root-mean-square residual of 0.28~dB over 6--30~dB; the edge separation $s = 37.1$~dB predicts through Fig.~\ref{fig:theory}(b) a plateau width of 17.1~dB and a deficit of 0.4~dB, against the measured 17.0~dB, and the fitted model places the plateau at 8.6--25.7~dB, against the measured 8.9--25.9~dB. System~A therefore operates in the well-separated regime of Proposition~3, and its plateau level of $-19.0$~dB lies within 0.7~dB of its hardware constant. System~D exhibits the low-loss regime that also appears in the record systems: a key rate that stays near 5~kb/s while the channel loss changes by 14~dB shows that a rate ceiling, from the system's rate control (for example $\mu$ adjustment) or from detector or processing saturation, sets its SKR. Its fit gives $N_0 = -26.3$~dB, $L_s = 26.0$~dB and $L_n = 40.6$~dB with a residual of 0.33~dB; the edge separation $s = 14.6$~dB predicts a plateau width of 6.1~dB, against the measured 6.7~dB, and a deficit of 4.8~dB. The plateau of System~D at $-31.5$~dB is therefore a lower bound on its hardware constant, set by this ceiling, and the model recovers the constant itself. The 12.5~dB between the plateaus of Systems~A and~D thus decomposes into 8.0~dB of hardware difference, a 6.3-fold lower product of the five factors of (\ref{eq:budget}) for System~D, and 4.4~dB of additional saturation deficit, and the 24.9~dB between Systems~A and~C reflects the COW protocol structure together with the detector settings. System~A agrees with the published deployments of the same platform, at $-19.3$, $-19.7$ and $-19.4$~dB, within 0.7~dB, from the first published deployment in 2014 to these testbed measurements; with a channel-loss uncertainty of 0.35~dB in (\ref{eq:sigma}), the four values give $\chi^2 = 2.04$ for three degrees of freedom ($p = 0.56$), so the per-pulse efficiency of this platform is stable across deployments, partner organisations and a decade of hardware generations; all four published deployments include Toshiba co-authors, so the agreement demonstrates the reproducibility of the platform.

\begin{table*}[t]
\centering
\caption{1-dB NNKR Plateaus, Loss Sensitivity and Model Fits for the Testbed Systems and a State-of-the-Art Record}
\label{tab:testbed}
\footnotesize\setlength{\tabcolsep}{3.6pt}
\begin{tabular}{@{}llcccccccccc@{}}
\toprule
\textbf{System} & \textbf{Protocol} & \textbf{Plateau} & \textbf{Width} & \textbf{NNKR} & \textbf{CoV} & \textbf{NNKR slope} & \textbf{SKR slope} & \textbf{Loss dependence} & \textbf{Model $N_0$} & \textbf{Model $s$} & \textbf{Fit rms} \\
 & & (dB) & (dB) & (dB) & (\%) & (dB/dB) & (dB/dB) & \textbf{removed} (\%) & (dB) & (dB) & (dB) \\
\midrule
A (Toshiba, production) & BB84 & 8.9--25.9 & 17.0 & $-19.0$ & 6.9 & $+0.022$ & $-0.982$ & 97.8 & $-18.3$ & 37.1 & 0.28 \\
B (Toshiba, research) & BB84 & 13.0--19.2 & 6.2 & $-19.9$ & 10.1 & $+0.035$ & $-0.985$ & 96.4 & -- & -- & -- \\
D (IDQ ClavisXGR) & BB84 & 22.1--28.8 & 6.7 & $-31.5$ & 5.4 & $+0.017$ & $-1.009$ & 98.3 & $-26.3$ & 14.6 & 0.33 \\
C (IDQ Clavis3) & COW & 10.4--15.3 & 4.9 & $-43.9$ & 6.8 & $+0.025$ & $-1.002$ & 97.5 & -- & -- & -- \\
Fan-Yuan'26~\cite{FanYuan2026} & BB84 & 4.7--15.8$^a$ & 11.1$^a$ & $-13.6$ & 6.0 & $+0.049$ & $-0.951$ & 94.9 & -- & -- & -- \\
\bottomrule
\multicolumn{12}{@{}p{\textwidth}@{}}{\footnotesize IDQ: ID Quantique; COW: coherent one-way; CoV: coefficient of variation (standard deviation divided by mean) of NNKR across the plateau. Slopes: least-squares slope of NNKR or SKR in dB against channel loss in dB over the plateau; loss dependence removed $= 1 - |b_\mathrm{NNKR}|/|b_\mathrm{SKR}|$. Model columns: fit of (\ref{eq:full}) to the BB84 systems with continuous curves; $s = L_n - L_s$ is the edge separation of (\ref{eq:width}). The COW protocol follows a different key-rate formula, and the curve of System~B contains step changes of up to 3~dB. $^a$Upper edge set by the longest reported distance.}
\end{tabular}
\end{table*}

\subsection{Coexistence Measurements}
\label{sec:coex}

\begin{figure}[t]
\centering
\includegraphics[width=0.92\columnwidth]{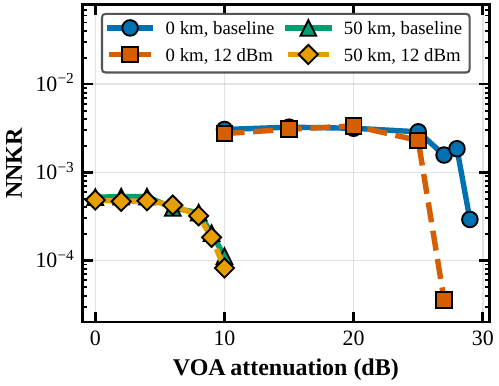}
\caption{Coexistence measurements on a Toshiba QKD system: NNKR against variable optical attenuator (VOA) attenuation at 0 and 50~km, at baseline and with 12~dBm classical injection.}
\label{fig:coex}
\end{figure}

Does either metric degrade when the QKD system shares its fibre with classical traffic? The coexistence set answers this question. Fig.~\ref{fig:coex} shows NNKR across the VOA sweep, at baseline and with 12~dBm of classical injection. At each fibre length, the two curves nearly coincide.

\begin{figure}[t]
\centering
\includegraphics[width=0.92\columnwidth]{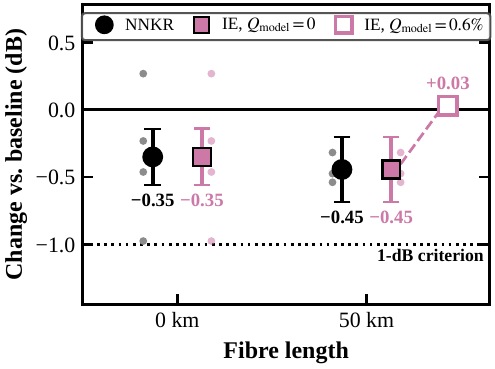}
\caption{Robustness of NNKR and IE in the coexistence environment of Fig.~\ref{fig:coex}. Each point is the change of a metric with 12~dBm of classical injection, relative to its baseline value. Only VOA settings on the baseline plateau are used: 10--25~dB at 0~km and 0--4~dB at 50~km. Small markers: individual settings. Large markers: means with standard errors. IE is shown for $\qber_\mathrm{model} = 0$ (filled) and, at 50~km, for $\qber_\mathrm{model} = 0.6$\% (open). Dotted line: the 1-dB plateau criterion.}
\label{fig:robust}
\end{figure}

Fig.~\ref{fig:robust} quantifies this agreement. For each VOA setting, it compares the metric with injection to its baseline value, $\Delta\nnkr = 10\log_{10}(\nnkr_\mathrm{12\,dBm}/\nnkr_\mathrm{base})$. Both values share the same clock rate, transmittance and link, including the ROADM path, so these factors cancel. Only settings inside the 1-dB plateau of the baseline are used: 10--25~dB at 0~km and 0--4~dB at 50~km.

NNKR changes by $-0.35\pm0.21$~dB at 0~km and by $-0.45\pm0.24$~dB at 50~km. The seven individual changes range from $-0.98$ to $+0.27$~dB, with a mean of $-0.39\pm0.16$~dB. Their scatter of 0.37~dB matches the 0.42-dB uncertainty of a single change. NNKR therefore stays within 0.5~dB of its standalone value. This is half of the 1-dB plateau criterion. It is also smaller than the 0.7~dB spread between the four determinations of this platform (Section~\ref{sec:lab}).

IE follows from the same data through (\ref{eq:ratio}). At baseline there is no classical light, so $\qber_\mathrm{model} = 0$ and $\ie = 2\,\nnkr$. With injection, the modelled noise lowers the reference $R_\mathrm{limit}$, and $\ie = 2\,\nnkr/[1-2H_2(\qber_\mathrm{model})]$. The SKR, clock rate and transmittance are common to both metrics. The change of IE is therefore
\begin{equation}
\begin{aligned}
\Delta\ie &= \Delta\nnkr - 10\log_{10}\big[1-2H_2(\qber_\mathrm{model})\big]\\
&\geq \Delta\nnkr .
\end{aligned}
\label{eq:dIE}
\end{equation}
Whatever the noise model, IE cannot degrade more than NNKR.

At 0~km, the classical light crosses no transmission fibre. The modelled Raman and four-wave-mixing noise vanish, so $\qber_\mathrm{model} \approx 0$. IE then changes exactly as NNKR does, by $-0.35\pm0.21$~dB. At 50~km, the correction term in (\ref{eq:dIE}) depends on the modelled noise. It is 0.10~dB at $\qber_\mathrm{model} = 0.1$\%, 0.47~dB at the 0.6\% modelled for Gavignet'24 at 16.8~dBm over 50~km (Section~\ref{sec:lit}; IE/NNKR $= 2.23$) and 0.77~dB at 1\%. Up to the Gavignet'24 value, IE at 50~km changes by between $-0.45$ and $+0.03$~dB. Both metrics therefore keep their standalone values in the coexistence environment.

The classical light does affect the curves, but only beyond the plateau (Fig.~\ref{fig:coex}). At 0~km, the 12-dBm curve leaves its plateau between 20 and 25~dB. The baseline curve leaves it between 25 and 27~dB. The lowest change in Fig.~\ref{fig:robust}, $-0.98$~dB at 25~dB, marks the start of this roll-off. At 27~dB, NNKR with injection lies 16.4~dB below baseline. At 50~km, the two curves separate by 1.35~dB at the last VOA setting, 10~dB.

The model of Section~\ref{sec:model} explains this behaviour. Classical injection adds noise detections at a rate that does not fall with the VOA attenuation. This raises $p_n$ in (\ref{eq:full}). On the plateau, signal detections outnumber noise detections by much more than $\rho_1$. The noise correction $c_n/\rho$ in (\ref{eq:cn}) therefore stays small, and NNKR stays at the hardware constant $N_0$ of Proposition~1, which has no noise term. As the attenuation grows, the ratio $\rho$ of signal to noise detections falls. With injection, it reaches $\rho_1$ at lower loss. The upper plateau edge, $L_\mathrm{hi} = L_n - 10\log_{10}\rho_1$ (Proposition~3), therefore moves to lower loss.

In short, coexistence shortens the plateau but does not change its level. Under real classical load, this production system operates in the same signal-dominated regime as the published experiments. On the plateau, NNKR and IE both report the hardware.

\section{Conclusion}
\label{sec:conclusion}

We have proposed two metrics for benchmarking QKD hardware across operating conditions and tested them on published and testbed data. The secret key rate factorises into the transmitter clock rate, the channel transmittance and a hardware term, the noise-normalised key rate, which needs four routinely reported parameters and reads as a decibel penalty budget relative to ideal BB84 at $-3$~dB. The implementation efficiency adds the coexistence noise of the channel to the reference, and for BB84 the ratio IE/NNKR depends only on the modelled noise error rate. An analytic key-rate model gives the hardware constant in closed form, derives the edges of the flat range from the product of detector noise count rate and dead time, and reproduces the measured curves with residuals of 0.28--0.33~dB. An objective 1~dB plateau criterion turns the metric into a measurement procedure, and four data sets, four consistency checks, a $\chi^2$ test of cross-laboratory agreement and an explicit uncertainty analysis substantiate it. Across ten coexistence experiments, IE places three deployments of one platform---across different fibres, bands and traffic loads---within a factor of 1.09, while separating the deployment configured for 16.8~dBm of classical power by 3.8~dB. In contrast, the CE metric spreads the same systems by a factor of 1470. On a multi-vendor testbed, four systems show a 1~dB plateau, and the production system holds the 1~dB plateau across 17~dB of attenuation with a 6.9\% coefficient of variation, at a level within 0.7~dB of the three earlier published deployments. In a coexistence environment, NNKR and IE of a production system of the same platform stay within 0.5~dB of their standalone values at 0 and 50~km. The consistent metric values indicate that the Raman and other impacts of coexistence in these experiments are negligible, and that the differences in measured key rate between them come from the QKD operating conditions, such as the system loss. The NNKR metric also provides a clear measure of the performance of different QKD systems relative to ideal limits. The framework thereby provides a vendor-neutral metric that can measure hardware and implementation evolution that is independent of propagation and can reveal performance differences for various operating conditions such as coexistence with classical communication signals.

\section*{Data Availability}
The testbed characterisation and coexistence measurements are available at https://doi.org/10.5281/zenodo.21792844. The figure data and analysis scripts are available from the corresponding author on reasonable request.

\bibliographystyle{IEEEtran}
\bibliography{references}

\end{document}